
\magnification=\magstep1
\vbadness=10000
\parskip=\baselineskip
\parindent=10pt
\centerline{\bf MATTERS OF GRAVITY}
\bigskip
\bigskip
\line{Number 3 \hfill Spring 1994}
\bigskip
\bigskip
\bigskip
\centerline{\bf Table of Contents}
\bigskip
\hbox to 6truein{Editorial {\dotfill} 2}
\hbox to 6truein{Correspondents {\dotfill} 2}
\bigskip
\hbox to 6truein{\bf Gravity news:\hfill}
\hbox to 6truein{Open Letter to gravitational physicists, Beverly
Berger   {\dotfill} 3}
\hbox to 6truein{A Missouri relativist in King Gustav's Court, Clifford
Will   {\dotfill} 6}
\hbox to 6truein{Gary Horowitz wins the Xanthopoulos award, Abhay Ashtekar
  {\dotfill} 9}
\bigskip
\hbox to 6truein{\bf Research briefs:\hfill}
\hbox to 6truein{Gamma-ray bursts and their possible cosmological
implications, Peter Meszaros  {\dotfill} 12}
\hbox to 6truein{Current activity and results in laboratory gravity,
Riley Newman {\dotfill} 15}
\hbox to 6truein{Update on representations of quantum gravity, Donald Marolf
{\dotfill} 19}
\hbox to 6truein{Ligo project report: December 1993, Rochus E. Vogt
{\dotfill} 23}
\hbox to 6truein{Dark matter or new gravity?, Richard Hammond
{\dotfill} 25}
\bigskip
\hbox to 6truein{\bf Conference Reports:\hfill}
\hbox to 6truein{Gravitational waves from coalescing compact binaries,
Curt Cutler
{\dotfill} 28}
\hbox to 6truein{Mach's principle: from Newton's bucket to quantum
gravity, Dieter Brill
{\dotfill} 31}
\hbox to 6truein{Cornelius Lanczos international centenary conference,
David Brown
{\dotfill} 33}
\hbox to 6truein{Third Midwest relativity conference, David Garfinkle
{\dotfill} 36}
\bigskip
\bigskip
\bigskip
\bigskip
\leftline{\bf Editor:}

\medskip
\leftline{Jorge Pullin}
\smallskip
\leftline{Center for Gravitational Physics and Geometry}
\leftline{The Pennsylvania State University}
\leftline{University Park, PA 16802-6300}
\smallskip
\leftline{Fax: (814)863-9608}
\leftline{Phone (814)863-9597}
\leftline{Internet: pullin@phys.psu.edu}

\vfill
\eject

\centerline{\bf Editorial}

Well, this newsletter is growing into its third year and third number
with a lot of strength. In fact, maybe too much strength. Twelve
articles and 37 (!)  pages. In this number, apart from the
"traditional" research briefs and conference reports we also bring
some news for the community, therefore starting to fulfill the
original promise of bringing the gravity/relativity community closer
together.

As usual I am open to suggestions, criticisms and proposals for articles
for the next issue, due September 1st.  Many thanks to the authors and
the correspondents who made this issue possible.

If everything goes well this newsletter should be available in the
gr-qc Los Alamos bulletin board under number gr-qc/9402002. To
retrieve it send email to gr-qc@xxx.lanl.gov (or gr-qc@sissa.infn.it
in Europe) with Subject: get gr-qc/9402002 (issue 2 is available as
gr-qc/9309003). Or email me. Have fun.

\medskip
\leftline{Jorge Pullin}

\bigskip
\bigskip
\centerline{\bf Correspondents}
\medskip

\parskip=2pt
\item{1.} John Friedman and Kip Thorne: Relativistic Astrophysics,
\item{2.} Jim Hartle: Quantum Cosmology and Related Topics
\item{3.} Gary Horowitz: Interface with Mathematical High Energy Physics,
    including String Theory
\item{4.} Richard Isaacson: News from NSF
\item{5.} Richard Matzner: Numerical Relativity
\item{6.} Abhay Ashtekar and Ted Newman: Mathematical Relativity
\item{7.} Bernie Schutz: News From Europe
\item{8.} Lee Smolin: Quantum Gravity
\item{9.} Cliff Will: Confrontation of Theory with Experiment
\item{10.} Peter Bender: Space Experiments
\item{11.} Riley Newman: Laboratory Experiments
\item{12.} Peter Michelson: Resonant Mass Gravitational Wave Detectors
\item{13.} Robbie Vogt: LIGO Project
\parskip=\baselineskip

\vfill
\eject

\centerline{\bf Open letter to gravitational physicists}
\medskip
\centerline{Beverly Berger, Oakland University}
\bigskip
\bigskip

\noindent I believe the time has come to create a Topical Group (TG)
in Gravitation within the American Physical Society (APS).
\medskip

\noindent Substructures of the APS include Divisions (e.g.\ Particles
and Fields, Nuclear Physics, Astrophysics), TG's (e.g.\ Metrology,
Laser Science), and Fora (e.g.\ Physics and Society).  The APS
constitution states
``If at least two hundred members wish to advance and diffuse the
knowledge of a specific subject or subfield of physics, they may
petition the Council [of the APS] to establish a Topical Group.''  If
the membership of the TG becomes sufficiently large, the TG can then
become a Division.
\medskip

\noindent Although the existing divisions of Particles and Fields and
Astrophysics have interests which include various aspects of
gravitational physics, their primary concerns lie elsewhere.  There are
many reasons to develop a TG at this time.  Such a group would allow us
to define and promote our interests and enhance our visibility within
the larger community of physicists.  The construction of LIGO and the
Grand Challenge Supercomputing Project on the two black hole problem are
only two examples of the significant developments in our field that
render such advocacy essential.
\medskip

\noindent There are several other advantages to the formation of the TG.
As a subgroup of APS, we would have access to APS public relations
activities
(e.g.\ press releases, lobbying) that might be marshaled for our benefit.
Some of our local meetings and workshops (e.g.\ the Pacific Coast or
Midwest Meetings) could become TG meetings with access to APS publicity
and possibly support.  Members of the TG could organize sessions at APS
general meetings. The TG would be a distinct entity through which to
support the positions and activities of the International GRG Society.
(Many countries already have separate national general relativity
societies.)  Finally, members of the TG can nominate candidates to
become APS Fellows.
\medskip

\noindent The procedures to form the TG are the following:  As quoted
above, a petition signed by 200 APS members requesting formation of
the TG must be received by the Executive Council of the APS.  (The
signed petitions must be actual hard-copy rather than electronic mail.)
The council, in consultation with existing APS divisions and TG's may
either accept or reject the petition.  If the signatures are obtained
and the petition is accepted, an organizational structure including TG
officers and bylaws must be put into place.  At that point, the TG
comes into existence.  Any APS member may join the TG for an additional
\$5 per year.
\medskip

\noindent The APS is open to physicists of any nationality for an annual
dues of approximately \$80 per year (less for students, postdocs, and
retirees).  One benefit of APS membership is the opportunity to subscribe
to some journals at a reduced rate.  The list includes {\it Physical
Review}, {\it Physical Review Letters}, {\it Journal of Mathematical
Physics}, and {\it Classical and Quantum Gravity} among others.
\medskip

\noindent If you agree with the need to form a TG in gravitation and
are an APS member please print out the attached petition, sign it, urge
your colleagues who are APS members to sign it, and mail the completed
petition to me.  I also invite 10-20 public spirited theoretical and
experimental relativists who are APS members to act as an {\it ad hoc}
organizing committee.  The function of this committee would be to
generate the bylaws of the organizational structure required for the
TG and to serve as an initial nominating committee for TG officers.
\medskip

\noindent I welcome your comments and questions.
\medskip

\noindent Sincerely,

\noindent Beverly K. Berger
\medskip

\noindent Physics Department

\noindent Oakland University

\noindent Rochester, MI  48309

\noindent (810)370-3422, (313)971-0345

\noindent berger@vela.acs.oakland.edu

\eject

\centerline{PETITION TO THE COUNCIL OF THE AMERICAN PHYSICAL SOCIETY}
\medskip

\noindent {We, the undersigned members of the American Physical
Society,
petition the Council of the American Physical Society to establish a
Topical Group in Gravitation.  Areas of interest to the proposed Topical
Group include, but are not limited to, experiments and observations
related to the detection and interpretation of gravitational waves,
experimental tests of gravitational theories, computational general
relativity, relativistic astrophysics, solutions to Einstein's equations
and their properties, alternative theories of gravity, classical and
quantum cosmology, and quantum gravity.  The purpose of the Topical
Group is to provide a unified forum for these areas of current research
which now span several Divisions of the Society.}
\medskip
\medskip

\centerline{Signature\hskip 3cm Name (Printed) \hskip 3cm Affiliation}
\bigskip
\bigskip
\bigskip
\hrule
\bigskip
\bigskip
\hrule
\bigskip
\bigskip
\hrule
\bigskip
\bigskip
\hrule
\bigskip
\bigskip
\hrule
\bigskip
\bigskip
\hrule
\bigskip
\bigskip
\hrule
\bigskip
\bigskip
\hrule
\bigskip
\bigskip
\hrule

\bigskip

\medskip

\noindent Return to:

\centerline{Beverly K. Berger}

\centerline{Physics Department}

\centerline{Oakland University}

\centerline{Rochester, MI  48309}

\vfill
\eject

\centerline{\bf A Missouri Relativist in King Gustav's Court}
\medskip
\centerline{Clifford Will, Washington University, St. Louis}
\bigskip
\bigskip
\medskip

The award of the 1993 Nobel Prize in Physics to Joseph Taylor and
Russell Hulse of Princeton University for their discovery of the
binary pulsar PSR 1913+16 puts a welcome seal of approval
on general relativity.  It was a long time
coming.

Einstein's general theory
has revolutionized our view of space and time
and the universe.  Yet the Nobel prize has never, until now, been
awarded for work so directly related to general relativity.
Even Einstein's Nobel in 1921 was for his work on the photoelectric
effect, not for relativity.
Part of the problem is that theory is difficult to test,
and the Nobel committees historically prefer to reward
work that has had experimental
confirmation.

Until 1974, the solar system provided the principal testing ground for
GR.  However, the discovery of the binary pulsar in the summer 1974
[1] showed that certain kinds of distant astronomical systems may also
provide precision laboratories for testing general relativity.  The
system consists of a 59 ms period pulsar in an eight-hour orbit with a
companion that has not been seen directly, but that is generally
believed to be a ``dead'' neutron star.  The unexpected stability of
the pulsar ``clock'' and the cleanliness of the orbit allowed Hulse
and Taylor and later co-workers to determine the parameters of the
system to extraordinary accuracy.  Furthermore, the system is highly
relativistic ($v_{\rm orbit} /c \approx 10^{-3}$).  Observation of the
relativistic periastron advance ($4^{\rm o}.22663 \pm 0^{\rm
o}.00002\, {\rm yr}^{-1}$), and of equivalence-principle effects on
pulse arrival times (gravitational redshift, time dilation -- 0.07\%
accuracy) can be used, assuming that general relativity is correct, to
determine the pulsar and companion masses, with the result $m_p =
1.4411 \pm 0.0007 M_\odot$ and $m_c=1.3873\pm0.0007 M_\odot$ .  The
measurement of the rate of decrease of the orbital period in 1979 [2]
gave the first evidence for the effects of gravitational radiation
damping.  With the measured orbital elements and the two masses, the
general relativistic quadrupole formula predicts the damping rate
$dP/dt=-2.40243 \pm 0.00005 \times 10^{-12}$.  The observations are
now better than 0.5 percent in accuracy, with $dP/dt_{\rm observed} =
-(2.408 \pm 0.011) \times 10^{-12}$, agreeing completely with the
prediction [3].  This verifies the existence of gravitational waves,
its quadrupole character, and the validity of the quadrupole formula
of GR.

Forty binary radio pulsars are now known.  Two of these, PSR 1534+12
in our galaxy, and PSR 2127+11C in the globular cluster M15, are
particularly promising as relativity laboratories.  Because of its
high timing accuracy and its proximity to the Earth, 1534+12 is
expected to yield an even more accurate determination of $dP/dt$ than
did 1913+16 [3].

In addition to verifying the existence of gravitational radiation,
binary pulsars provide ``strong-field'' tests of general relativity,
in contrast to the solar-system ``weak-field'' tests, in the following
sense.  Since such systems contain at least one, and probably two
neutron stars, the bodies contain strongly relativistic internal
gravitational fields.  In most alternative theories of gravity (but
{\bf not} GR), the motion of compact objects is affected by their
internal structure (violation of the Strong Equivalence Principle); in
addition, most theories predict ``dipole'' gravitational radiation in
addition to the quadrupole part, whose source is the difference in
internal gravitational binding energies of the two stars.  Because of
these two phenomena, many alternative theories of gravity, which
otherwise might agree with solar-system observations, can be strongly
tested by binary pulsar systems (for review and references see [4]).

Binary pulsars are also important as a
foundation for gravitational-wave observatories such as LIGO and
VIRGO.  Because their lifetimes against coalescence induced by
gravitational radiation reaction
are short compared to the age of the galaxy, one can expect
that coalescing binary neutron stars are happening today.  Based on this
fact, several estimates have been made of the coalescence rate,
leading to numbers of the order of 3 per year with a detectable signal
strain of $10^{-21}$ in a volume of the
universe 200 Mpc on a side.

The actual award ceremony took place in Stockholm during the week of
December 7-11.  I and my wife had the honor and privilege of attending
the festivities as guests of the Nobel Committee in recognition of my
role as a ``special referee'' in helping them make the decision,
formulate the citation, and so on (Alvaro de R\'ujula of CERN was the
other physicist so honored).  The experience was fantastic, and I
highly recommend it to everyone (the only assured way of being
invited, however, is to win the prize)!  The Laureates, families and
guests stayed at the Grand Hotel of Stockholm, the kind of place that,
if you have to ask the price, you probably don't belong there.

The activities kicked off on Tuesday the 7th with a press conference
and reception for the Physics, Chemistry and Economics winners at the
Swedish Academy of Sciences.  Not surprisingly, most of the press
questions were for the Economists and for the Chemists (who worked on
aspects of DNA replication immortalized in {\it Jurassic Park}); GR
still seems to be a mystery to most of the press.  Wednesday morning,
Hulse and Taylor gave the Nobel lectures.  Hulse, who was Taylor's
University of Massachusetts graduate student during that fateful
summer, described the actual steps involved in the discovery of the
binary pulsar.  Ironically, following his Ph.D. and a post-doctoral
stint, Hulse left radio astronomy and now does plasma physics at the
fusion laboratory at Princeton.  Taylor, now the James S. McDonnell
Professor of Physics at Princeton, related the techniques by which the
system is used for tests of relativistic gravity and of the existence
of gravitational waves.  Other events included a special Nobel concert
by the Stockholm Philharmonic, and a big reception hosted by the Nobel
Foundation at the ornate Great Hall of the Royal Swedish Academy.

By far the highlight of the week, however, was the award ceremony at
the Concert Hall and the banquet at the Stockholm City Hall.  Required
dress was white tie and tails for the men and ball gowns for the
women.  For the Physics Prize, the citation read by Professor Carl
Nordling of the Nobel Committee emphasized the uniqueness of PSR
1913+16 as the first radio binary pulsar, and its importance as a new
laboratory for testing general relativity and revealing the existence
of gravitational waves.  Then Hulse and Taylor stepped forward to
receive their prizes from the King.  Following the awards, the
participants were limousined or bused to the City Hall for the
banquet, held in the ``Blue Hall''.  (Some readers will recall that
this was the site of the buffet dinner at GR 11 in 1986.  Joe Taylor
was an invited speaker at GR 11, and participated in that buffet
dinner -- kind of a rehearsal for the real thing).  Counting the
Laureates in Physiology/Medicine and Literature (the author Toni
Morrison -- {\it Beloved}, {\it Jazz}), their families and guests, the
members of the various Nordic Academies and their guests, the list
came to an intimate 1300 people.  A 60-page book, complete with
fold-out map, gave the seating plan (Hulse sat next to the Queen).
One nice tradition is the inclusion of about 250 university students
in the festivities, each wearing a cap distinctive of his or her
institution.  The four-course meal, served on special gold-trimmed
Nobel table settings created for the 1991 Jubilee of the Prize, was
served with military precision by some 130 waiters.  The wine was
served by choral groups who sang Swedish songs as they poured.
Although you might think such an affair would be formal and stuffy, it
was far from it.  The thing was a total blast.  The dance following
was lively, although the orchestra must have experienced severe time
dilation some time in the past, for it had not yet reached the 1960's.
Some of the post-dance parties went on until 5:00 in the morning.  The
final event for the Physics people was a panel discussion at the Royal
Institute of Technology, with Taylor, Hulse, de R\'ujula, Thibault
Damour and me as panelists.

Taylor graciously invited to the festivities as one of his personal
guests Jocelyn Bell-Burnell, the co-discoverer of pulsars in 1967, who
most astrophysicists feel was wrongly overlooked for the 1974 Nobel
Prize, shared by her Ph.D. adviser and co-discoverer Anthony Hewish
and radio astronomer Sir Martin Ryle.  This year's situation was quite
similar: Hulse was Taylor's graduate student, doing the day-to-day
work at the Arecibo radio telescope in Puerto Rico, while Taylor was
on campus in Amherst.  Hulse actually made and confirmed the
discovery, and then he and Taylor together made the initial accurate
orbit determinations and measured the periastron advance (for a
detailed account of the discovery, see [5]).  Subsequent work by
Taylor and other co-workers enabled a determination of the orbital
damping.  This time, however, the Nobel Committee properly gave the
prize to both men.

This newsletter seems an appropriate place, on behalf of the
U.S. general relativity community, to salute Russell Hulse and Joseph
Taylor on the award of the 1993 Nobel Prize in Physics.

\parskip=3pt
\medskip
\item{[1]} R. A. Hulse and J. H. Taylor, {\it Astrophys. J. Lett.}
{\bf 195}, L51 (1975)
\item{[2]} J. H. Taylor, L. A. Fowler and P. M. McCulloch, {\it
Nature} {\bf 277}, 437 (1979)
\item{[3]} J. H. Taylor, A. Wolszczan, T. Damour and J. M. Weisberg
{\it Nature} {\bf 355}, 132 (1992)
\item{[4]} C. M. Will, {\it Theory and
Experiment in Gravitational Physics} Revised Edition, (Cambridge
University Press, Cambridge, 1993)
\item{[5]} C. M. Will, {\it Was Einstein Right?}, 2nd Edition (Basic
Books, New York, 1993), Chapter 10
\parskip=\baselineskip

\vfill
\eject

\centerline{\bf Gary Horowitz wins the Xanthopoulos Award}
\medskip
\centerline{Abhay Ashtekar, Penn State University}
\bigskip
\bigskip
\medskip

The second international Basilis Xanthopoulos award in Relativity and
Cosmology was awarded to {\it Professor Gary T. Horowitz} of the
University of California at Santa Barbara for his wide ranging
contributions to the mathematical aspects of gravitational physics.
The citation singled out, in particular, Horowitz's work on positive
energy theorems, his influential ideas on global problems in general
relativity and his numerous papers which have illuminated the global
properties of space-time geometry in string theory. The award honors
the memory of the well-known relativist Basilis Xanthopoulos who was
shot to death while he was giving a seminar in Crete on November 27th,
1990. This unspeakable act of violence ended without forewarning a
life of joy and energy, rich in accomplishment and promise.

The award ceremony took place during the reception at the Lanczos
Centennial conference at Raleigh, North Carolina on Sunday, December
12th. Professor James York of the University of North Carolina chaired
the session. He introduced Professor Sotirios Persides of the
University of Thessaloniki who had flown from Greece specially for the
occasion. Persides was in the audience during the seminar on that
fateful day in November 1990 and thanks to his extraordinary personal
courage and self-sacrifice, many lives were saved. He himself was
badly wounded and it is only now that he is close to a full recovery.
Persides recalled the origin of the award.  Since the readers of this
newsletter may not all be aware of this, we will reproduce his remarks
in full:

{\sl I am here today on behalf of the Foundation for Research and
Technology Hellas (FORTH) to say a few words about the Basilis
Xanthopoulos award and about our beloved colleague.

The seminar on algebraic computing has started that evening of
November 27, 1990, at the University of Crete, the Greek inland.
Basilis Xanthopoulos was lecturing on the blackboard and about 25
professors and graduate students were listening.  Suddenly, a deranged
gunman burst into the room and opened fire.  Basilis Xanthopoulos, at
his 39th year of life, and Stefanos Pnevmatikos, at his 33rd, two
young professors, internationally known scientists, and dear friends
of ours were brutally murdered.  Another professor and a female
student were injured.  Personally, in a desperate act of defense, I
was severely wounded.  A distraught mind and the availability of guns
have brought death, distraction and unbearable pain.  Basilis and
Stefanos, two teachers of scientific truth, were lost at a moment
when their studies and academic work had started to blossom.

Basilis Xanthopoulos was born in Drama, Greece, in 1951.  He was the
best student I had at the University of Thessaloniki where he received
his B.S.  in Mathematics in 1973.  After his military service, he
followed graduate studies at the Physics Department of the University
of Chicago, where he became a teaching assistant and later a research
assistant.  He received his M.S. in 1976 and his Ph.D. in 1978 with
Robert Geroch as his advisor.  He continued as a visiting assistant
professor at the Physics Department of Montana State University and
Research Associate at Syracuse University.  In December 1979, he
returned as chief assistant at the Astronomy Department of the
University of Thessaloniki.  Finally, in 1982, he was appointed as an
assistant professor at the Physics Department, University of Crete,
where he became associate professor the next year and professor in
1987.  During his career, he held many positions in various
universities and research centers.  From September 1987, he was
Chairman of the Physics Department of the University of Crete.

Basilis Xanthopoulos had a passionate love for science.  His main
field of research was theoretical general relativity.  He published
about 60 original research papers in scientific journals trying to
unravel the complexities of mathematical gravity.  He worked on the
asymptotic structure, exact solutions, perturbations of black holes,
gravitational waves, Yang-Mills fields, formation of singularities and
many other topics of intense mathematical nature in the framework of
the Einstein and Einstein-Maxwell geometry of space-time.  These
articles appeared in distinguished international journals.  He
collaborated with many well known scientists as R. Geroch, A.
Asthekar, S. Chandrasekhar, C.  Hoenselaers, W. Kinnersley, V. Ferrari
and others.  He gave lectures on his research work in many
universities, research institutes and scientific conferences.

As a teacher Basilis was also exceptional.  He transmitted knowledge
to his audience in an informal and friendly way, clearly stating the
known, the assumptions, the reasoning and the conclusions, captivating
specially the young.  He organized seminars, undergraduate and
graduate courses, summer schools, conferences.  He worked hard for
many years for science, for the students, for the University of Crete.
He was always willing to help his colleagues and his students.  He was
always cheerful and philosophical in front of the difficulties of
life.  But, alas!  As Professor Chandrasekhar, his teacher,
collaborator and friend has noted, the violent termination of Basilis'
life, on the verge of a most promising career, was an immense tragedy
for science and for the University of Crete.  I believe it is a severe
loss for Greece, too, specially since Basilis worked in general
relativity, a field where it is very difficult to find and train
talented people.

To honor his memory, the Foundation for Research and Technology-Hellas
has established an International Award for Relativity (carrying a
financial value of \$9,000) to be given every two years to a young
scientist (below 40) who has published outstanding work in
gravitational physics (preferably theoretical).  An international
committee of distinguished scientists with S. Chandrasekhar as
Chairman has awarded the first award in 1991 to Demetrios
Christodoulou of the Courant Institute.  Today, we are here for the
presentation of the second Basilis Xanthopoulos Award.}

The committee which selected the 1993 winner of the award consisted of
Professors Subrahmanyan Chandrasekhar (Chair), George Contopoulos,
Roger Penrose, Kip Thorne and myself. Since Professor Chandrasekhar
was out of the country at the time, I ---being the committee member
who was most closely associated with Xanthopoulos--- had the pleasure
of presenting the award. In his acceptance speech, Horowitz recounted
the story of a delightful collaboration he had with Xanthopoulos. As
graduate students, one Christmas the two of them wrote a paper on how
Santa Claus draws on a multitude of relativistic effects --including
the extraction of rotational energy from a Kerr black hole at the
North Pole-- to carry out his many wonderful tasks.  The story was
picked up by many newspapers and, according to Horowitz, is the most
widely known work of his!

\vfill
\eject

\centerline{\bf Gamma-ray Bursts and their Possible Cosmological Implications}
\medskip
\centerline{Peter Meszaros, Penn State University}
\bigskip
\bigskip
\medskip

Gamma-ray bursts (GRBs) are brief gamma-ray flashes detected with
space-based detectors in the range 0.1-100 MeV, with typical photon
fluxes of $10^{-2}- 10^{2} {\rm cm}^{-2}$ s$^{-1}$ and durations of
$10^{-1}-10^3$ s. Their origin is clearly outside the solar system,
and more than a thousand events have been recorded so far.  Before
there was any firm evidence on the isotropy of classical gamma-ray
bursts, the most plausible interpretations involved magnetospheric
events on neutron stars (NS) within our Galaxy. However, the
remarkable isotropy of these events discovered within the last two
years by the BATSE experiment on the NASA Compton Gamma Ray
Observatory (together with the `flatter than Newtonian' counts)
clearly shifts the odds substantially in favor of a cosmological
interpretation.

In principle, the isotropy could be interpreted in terms of either (1)
a cosmological distribution similar to that of the distant galaxies
and clusters, i.e. hundreds of Mpc, (b) a distribution in an `extended
halo' of our galaxy, which is so large that the small dipole moment
associated with our off-center location is not noticeable
(i.e. greater than 50 Kpc), or (c) a `galactic disk' distribution,
where objects are sufficiently faint that they are detectable only out
to distances smaller than the width of the disk (less than a few Kpc).
Of these, the `galactic disk' model does have a semi-plausible energy
source: the fluences required are of the order of $10^{39}-10^{40}$
ergs, and these could be associated with brief, mildly super-Eddington
outflows from neutron stars undergoing a star-quake, comet impact or
brief accretion episode. However, it is difficult to explain a large
number of events (a few per day) occurring within a few Kpc. The
`extended halo' option is even more difficult, since it is hard to see
how such an extended halo would have formed or survived tidal
stripping by neighboring galaxies, the frequency of occurrence is not
much easier to explain, and the required fluences ($10^{41}-10^{43}$
ergs) are much larger than for typical brief events on neutron
stars. On the other hand, the `cosmological' interpretation does have
at least two rather plausible energy sources: either NS-NS (or
NS-black hole) binary mergers (e.g. binary pulsars merging under the
effect of gravitational wave energy losses), or else `failed
supernova' events (where a star undergoes core collapse to a NS but
with much reduced optical display). Either of these should occur with
a frequency of $10^{-5}$ per galaxy per year, and produces
$10^{50}-10^{51}$ ergs, visible out to redshifts $z$ of order unity,
so that the typical frequency and fluence is easily explained.

Irrespective of the typical distance (but particularly in cosmological
models), the energy density in a GRB event is so large that an
optically thick photon/$e^{\pm}$ fireball is expected to form, which
will expand carrying with itself some fraction of baryons. A question
of major interest is how such an event can lead to an optically thin,
non-thermal (generally broken power-law type) gamma-ray spectrum.  The
radiation burst escaping when the fireball becomes optically thin is
much too brief, has a quasi-thermal spectrum, and carries only a very
small fraction of the total energy (most of which ends up as kinetic
energy of the expanding baryons). However, all of this energy will be
reconverted into radiation when the baryons are decelerated by the
external medium. This results in a blast wave moving into the external
gas, and a reverse shock moving into the baryonic ejecta, both of
which are optically thin by the time deceleration occurs. If the
baryon loading is small, the expansion is relativistic, and both the
blast and reverse shocks will be strong, leading to acceleration of
particles with a power-law distribution of relativistic
energies. These can radiate efficiently even if the magnetic fields in
the shocked regions (which may be amplified by turbulence) are well
below equipartition. The resulting burst of gamma-rays has a total
energy and time-delayed duration of the order of that observed. The
synchrotron-inverse Compton spectrum produced in the shocks is also in
rough agreement with observations.  One can also predict the optical,
X-ray and TeV spectrum expected from such bursts, which in principle
could also arise in galactic models. The calculations indicate that
cosmological models generally produce much lower optical and X-ray
fluxes, in agreement with current non-detection upper limits, whereas
galactic models (except for short ones of less than about 1 s) would
predict X-ray fluences comparable to gamma-ray ones, and conspicuous
optical fluences, in disagreement with observations.  Even if an
adequate number of galactic sources were available, the predicted
spectra would be acceptable only if the duration is determined by an
intrinsic time scale, rather than by the more straightforward
relativistic time-delay arguments.

A straightforward prediction of cosmological models is that, if GRB
are "standard candles", one would expect the weaker fluence bursts
(which presumably are farther) to have longer durations due to
cosmological time-dilation. Such an effect has been recently reported,
both in preprint and New York Times form.  However, the duration of
the burst can depend on intrinsic properties of the event, such as the
total energy, so that this cosmological signature may be masked by
details of the source physics, which are model-dependent.  If
gamma-ray bursts are indeed cosmological, one very interesting
consequence would be that gravitational wave bursts of energy
comparable to a solar rest mass should occur at the same time as the
GRB events. These would be detectable at the rate of several per year
with coincidence measurements from two advanced versions of the
proposed LIGO or VIRGO detectors, at frequencies of $10^2-10^3$
Hz. Such measurements might also be able to distinguish between failed
supernova events, NS-NS or NS-black hole mergers, through their wave
profile. It will probably be hard to obtain any reliable information
concerning cosmological parameters, such as the closure parameter or
the Hubble constant, due to the smearing effects introduced by the
luminosity or density evolution of the bursts over cosmological
time scales.  Similarly, potential information about large-scale
structures (such as voids or superclusters of galaxies) will probably
be diluted beyond recognition by evolutionary effects. However, one
may obtain valuable information concerning early star formation,
through limits on the typical redshift derived from the counts of
events as a function of the fluence, and it may be possible to derive
limits on the GRB luminosity distribution.
\vfil
\eject

References:
\item{}
Meegan, L.A., et.al., 1992, Nature, 335, 143.
\item{}
Meszaros, P. and Rees, M.J., 1992a, Ap.J., 397, 570.
\item{}
Meszaros, P. and Rees, M.J., 1992b, M.N.R.A.S., 257, 29P.
\item{}
Meszaros, P. and Rees, M.J., 1993, Ap.J., 405, 278
\item{}
Meszaros, P., Laguna, P. and Rees, M.J., 1993, Ap.J., 215, 181.
\item{}
Narayan, R., Paczynski, B. and Piran, T., 1992, Ap.J.(Letters), 395, L83
\item{}
Paczynski, B., 1986, Ap.J.(Lett.), 308, L43
\item{}
Rees, M.J. and Meszaros, P., 1992, M.N.R.A.S., 258, 41P.
\item{}
Shemi, A. and Piran, T., 1990, Ap.J.(Lett.), 365, L55
\item{}
Woosley, S., 1993, Ap.J., 405, 273

\vfill
\eject

\centerline{\bf Current activity and results in laboratory gravity
experiments}
\medskip
\centerline{Riley Newman,  University of California, Irvine}
\bigskip
\bigskip
\medskip

\noindent {\bf G.}  At least five groups are actively
measuring G (still by far the least well-known fundamental
constant):  Luther at Los Alamos, Karagioz in Russia, Schurr at
the University of Wuppertal, Germany, Michaelis at the PTB (German
NIST equivalent), and Luo Jun in China, with their respective
collaborators.  The Wuppertal group uses a novel technique: two
suspended masses forming the walls of a microwave Fabry-Perot
resonator act as a gradiometer in response to a source mass.
Schurr reports that they have determined G to 220 ppm, with a
result in good agreement with the \lq\lq CODATA" value (whose
assigned uncertainty is 128 ppm); an improved experiment is in the
works.  The PTB group uses a mercury-supported balance, developed
by de Boer and collaborators; their experiment is recently
completed and they will soon publish in Metrologia a surprising
result:  $G = 6.71540 ~\pm ~0.00056 \cdot 10^{-11} N m^2 kg^{-2}$
(83 ppm uncertainty); about 1.006 times the currently accepted
(CODATA) value!  The group reports it could find no error in their
work in spite of a long period of searching.  Karagioz's results
are published [1], but not yet available in English translation
and I have lost e-mail contact with Russia.  New methods for
measurement of G in the laboratory (with homogeneous fields) and
with satellites are under development in Russia, V.N. Melnikov
reports.  (Melnikov expressed an interest in collaboration and
joint projects with groups in the USA.)

\bigskip

\noindent {\bf Funny gravity/new forces/weak equivalence
principle.}  These topics are intertwined -- operationally, one
man's feeble \lq\lq fifth force'' could equally well be another's
\lq\lq anomalous gravity.''  An excellent review of this area as
of 1991 is given by Adelberger, Heckel, Stubbs and Rogers [2];  a
comprehensive index of relevant publications as of 1992 has been
compiled by Fischbach, Gillies, Krause, Schwan, and Talmadge [3].
Here I will informally give the flavor of recent results and
current activity.  I believe no one currently claims evidence for
anomalies.

\bigskip

\noindent {\bf Inverse square law tests.}  Anomalous forces with a
range between 1 meter and a few kilometers remain the most poorly
constrained. Assuming an effective potential
$- GM (1 + \alpha e^{-r/\lambda})/r$, ~$\alpha$ is constrained to
be less than about .002 for these ranges by the value of $G
{}~(6.677 ~\pm ~0.013 \cdot 10^{-11}$) determined from submarine
measurements through a 5-km-thick slab of sea water [4].  Zumberge
reports that his group is developing a deeply towed unmanned
gravity meter with which a much larger volume of sea can be
sampled over a longer time; he hopes in a couple of years to
repeat the G experiment with much higher precision.  Meanwhile,
lab experiments chip away at ($\alpha, \lambda$) parameter space
for increasing ranges.  Goodkind et al. [5] report an experiment
in which an attracting mass is placed at various heights below a
gravimeter based on a levitated superconducting ball;  their
result for mass separations from 0.4 to 1.4 m is consistent with
earlier constraints.

Direct tests of the inverse square law are extremely sensitive to
uncertainties in source mass distribution.  Two ingenious
approaches are being developed which largely circumvent this
problem:  1. Measure the Laplacian of the gravitational potential,
$\nabla^2 \Phi$, or:  2. Measure the gradient of this quantity.  A
non-vanishing value for either of these (in vacuum) is a signal of
inverse square law violation, independent of source mass
distribution.  The first approach is used by Paik at Maryland, who
continues to refine his three-axis superconducting gradiometer to
measure $\vec{\nabla} \cdot \vec{g}$ as a test of Gauss's law;
his experiment using a swinging 1.5 ton lead ball source mass,
reported last spring [6], places a $2 \sigma$ limit: $\alpha =
(0.9 ~\pm ~4.6) \times 10^{-4}$, at $\lambda = 1.5$ m, improving
by an order of magnitude on existing limits at this range.  Paik
is developing [6] a null source -- a shaped hollow cylinder to
surround his gradiometer -- with which he hopes to achieve
sensitivity to $\alpha$ at a level of $10^{-6}$ at 0.2 m.  Using
geological-scale sources Paik sees a potential sensitivity to
$\alpha =  10^{-5}$ at 100 m.  The second approach to an inverse
square test insensitive to source mass detail -- measurement of
the gradient of the Laplacian -- is being pursued by Boynton at
the University of Washington [7].  Boynton's student Moore has
shown how to build a torsion balance whose (design) conventional
multipole moments vanish to high order, with correspondingly zero
coupling to fields whose Laplacian vanishes, but which has non-
zero moments of $r^3 Y_{\ell m}$ for $\ell = 1$ and $m = \pm~ 1$
-- moments which couple to the gradient of the Laplacian to
produce torques on the balance.  Boynton and collaborators plan
two experiments with this instrument -- one with a local source
mass designed to produce minimal gravity gradients (for a second
level of defense against spurious Newtonian signals), and one to
use a mountainside  as source.  They expect near term results from
the two experiments at a level $\alpha = 10^{-4}$, for ranges of
order 10 cm and 10 meters respectively, with a potential for much
greater sensitivity later with tighter fabrication tolerances.

The Wuppertal group is also pursuing $1/r^2$ tests using the
instrument described under G above.  No tower or borehole tests
are currently in the works, as far as I know.  In Japan, Hideo
Hanada concludes [8] that sensitivity to $\alpha$ on order $10^{-
6}$ for $\lambda < 10^5$ should be possible from gravity surveys
of the earth with a gravity gradiometer.

\bigskip

\noindent {\bf Composition dependence/weak equivalence principle.}
The big news here is dramatically improved limits on couplings to
N - Z (or B-2L), from controlled local source experiments.  Put in
terms of an inequality of the accelerations of a neutron and a
proton toward a neutron at a distance greater than one meter (and
assuming electrons behave normally), the best limit in 1991 was
about 185 ppm.  This has been reduced by the TIFR group in India
[9] to 120 ppm, in an experiment achieving an unprecedented
acceleration sensitivity of $10^{-13} cm/s^2.$  But the n-p
acceleration difference limit has now shrunk to 8 ppm, based on
preliminary results announced last summer [10] by the \lq\lq
E$\ddot{\rm o}$t-Wash'' group.  This group uses a stationary
torsion balance to search for differential acceleration of copper
and lead toward a rotating 3 ton uranium attracting mass.  As the
uranium extends to within 10 cm of the test masses, this
experiment not only dramatically improves constraints on anomalous
forces with range above 1 meter but also yields significant
constraints for smaller ranges than heretofore probed (down to 1
cm).

The rotating torsion balance of the E$\ddot{\rm o}$t-Wash group
has yielded improved equivalence principle test preliminary
results, using the earth's field as acceleration source [11]:

\medskip
	$\eta (Be,Cu) =  (- 4.0 ~\pm ~2.8) \times 10^{-12}$

	$\eta (Be,Al) =  (- 2.7 ~\pm ~3.4) \times 10^{-12}$
\medskip

Following a suggestion of Stubbs', the E$\ddot{\rm o}$t-Wash group
has used their instrument for an ingenious test of the
`gravitational' properties of dark matter, demonstrating that the
accelerations of the tested materials -- Be, Al, and Cu -- toward
dark matter [11] must be equal to about 3 parts/thousand ($2
\sigma$ limits).

Improved WEP/anomalous force experiments are currently being developed
by a number of groups.  The E$\ddot{\rm o}$t-Wash group is building a
completely new rotating torsion balance, with longer fiber.  Boynton's
group at UW is beginning operation of a new torsion balance at their
cliff-side \lq\lq Index'' site in Washington, and plans to develop a
new mountainside site at DOE's Hanford reservation -- overlooking one
of LIGO's sites.  At Irvine, we are testing a new rotatable torsion
balance, and are beginning a systematic test of the Q's of various
torsion fiber materials at 4.2K with an eye toward the low thermal
noise promised by cryogenic torsion balances.  At U. Mass Amherst,
Krotkov (of Roll, Krotkov, Dicke WEP test fame) has been working on a
torsion balance experiment testing for differential acceleration of
copper and polyethylene toward a modulated water source (pumped
storage reservoir).  The University of Bremen in Germany is setting up
an experiment using the drop tower of Bremen, with 110 m free fall --
their goal is a test of WEP with $10^{-12}$ accuracy.

\bigskip

\noindent {\bf Do antiprotons fall up}?  Adelberger argues
persuasively that in the usual scaler/vector picture of how a new
force could work, present limits on anomalous forces make it clear
that an antiproton should fall down.  But whether it {\bf does}
can't be known until one is actually dropped and watched.  Michael
Nieto reports that the Los Alamos-CERN collaboration hopes to
measure the acceleration of antiprotons in the earth's
gravitational field with a precision of better than 1\% by 1997.
Toward this end, the collaboration has already successfully
trapped a million antiprotons from a single accelerator pulse,
holding them for several minutes while cooling them to an energy
of a few eV.

\bigskip

\noindent {\bf Spin-dependent new-forces/gravity}?  I will report
on this area in a future issue of \lq\lq Matters of Gravity".


\item{}
[1] Izmerit. Technika, No. 10, p.3, (1993)

\item{}
[2] E.G. Adelberger et al., Ann. Rev. Nucl. Part. Sci. {\bf 41},
269 (1991)

\item{}
[3] E. Fischbach et al., Metrologia {\bf 29}, 213 (1992)

\item{}
[4] M.A. Zumberge et al., Phys. Rev. Lett. {\bf 67}, 3051 (1991)

\item{}
[5] J.M. Goodkind et al., Phys. Rev. D {\bf 47}, 1290 (1993)

\item{}
[6] M.V. Moody and H.J. Paik, Phys. Rev. Lett. {\bf 70}, 1195
(1992)

\item{}
[7] M.W. Moore et al.*

\item{}
[8] Hideo Hanada, Journal of the Geodetic Society of Japan {\bf
38}, 221

\item{}
[9] R. Cowsik et al., Proceedings of the XIIIth Moriond Workshop,
Switzerland, (1993)

\item{}
[10] J.H. Gundlach*

\item{}
[11] E.G. Adelberger*

\item{}
[12] G. Smith et al., Phys. Rev. Lett. {\bf 70}, 123 (1993)


*References 7,10, and 11 appear in the Proceedings of the
International Workshop on Experimental Gravitation, Nathiagali,
Pakistan (1993), to be published.

\vfill
\eject

\centerline{\bf Update on Representations for Quantum Gravity}
\medskip
\centerline{Donald Marolf, Penn State University}
\bigskip
\bigskip
\medskip

\def \ag { ${\cal A}/{\cal G}$ }
\def \ai { $\overline {{\cal A}/{\cal G}}$ }
\def \aim { \overline {{\cal A}/{\cal G}} }
\magnification = \magstep1
\baselineskip = 12pt

I would like to report on some recent
progress toward building representations for
canonical quantum gravity.  Two new perspectives have been found on
spaces that can be used as domain spaces of wavefunctions in a
quantum theory and I will describe each of these in turn.  The
first concerns the so-called ``connection representation" and the
second is a new ``extended loop representation."
Both provide excellent prospects for further advancement.

It will be good to begin with a reminder of what connection
representations are and what they have to do with quantum gravity.
Recall that one way to canonically quantize a system uses the
``configuration space representation" in which quantum states are
represented by functions on the configuration space of the system, or
on some closely related space.  Familiar examples are position
representations in one particle quantum mechanics and the
Schr\"odinger representation of free scalar quantum field theory
(QFT).  In the QFT case, the states are not just functions on the
classical configuration space of smooth fields but depend in an
essential way on {\it distributional} fields.  The situation can be
summarized as follows: The Fock representation of free scalar QFT is
unitarily equivalent to a representation in terms of functions on the
space ${\cal S}'$ of tempered distributions (distributional field
configurations) in ${\bf R}^3$ with an inner product given by
integration with respect to a Gaussian measure $\mu_G$.  Thus, the
Fock space is unitarily equivalent to the space $L^2({\cal
S}',\mu_G)$.  While the space ${\cal S}$ of smooth rapidly decreasing
fields is dense in ${\cal S}'$ and ${\cal S}'$ may be considered as a
completion of the classical configuration space, the smooth fields
actually belong to a set of measure zero with respect to the Gaussian
measure $\mu_G$.  Surprising as it seems, this means that the value of
the wave function on the classical configuration space has no effect
on the quantum state of the system.

Something similar is expected to be the case for quantum gravity and
a structure along these lines has been developed.  This takes place in the
context of the connection representation associated with the
so-called ``new variables" or ``Ashtekar variables." This formalism
describes gravity
in terms of a complexified $SO(3)$ connection and
a complex triad which are canonically conjugate.
Taking the connection to be the configuration
variable, the configuration space is just space ${\cal A}$
of smooth connections.

{}From the canonical perspective, such a description of gravity involves
three kinds of gauge symmetries:  complex $SO(3)$ rotations,
spatial diffeomorphisms, and Hamiltonian gauge transformations.  In
principle, we might like to work with a ``reduced configuration
space," which is the quotient of ${\cal A}$
by the action of these gauge groups.  However, this is not possible
for gravity because the Hamiltonian constraint does not map the
configuration space into itself but instead mixes coordinates and
momenta.  Thus, the
approach taken is to deal with each class of gauge transformations
separately.

The first step in this process is to consider the (partially)
reduced configuration space ${\cal A}/{\cal G}$ where ${\cal G}$ is the
space of complexified $SO(3)$ gauge transformations.
Suitable structures should be found on this space or
an appropriate completion that have nice properties under the action of
spatial diffeomorphisms.  This will allow a space of ``diffeomorphism
invariant wavefunctions" to be defined, on which the
Hamiltonian constraint is to be enforced.  Note that while
a Gaussian measure could be defined on the space ${\cal A}$ and this
space could be completed in direct analogy with the scalar QFT case, this is
based on the structure of ${\cal A}$ as a linear space -- a
structure that is not preserved by the action of ${\cal G}$.

Ashtekar and Isham [1] provided a completion
of \ag by considering a commutative C*-algebra
of traces of holonomy functions (Wilson loop functions). Their completion is
what is technically called the ``spectrum" of this algebra, which
is the completion of ${ \cal A}/{\cal G}$ in a particular topology due to
Gel'fand.  However, the holonomies of a connection will form
such a C*-algebra only if they are bounded functions; that is, only
if the gauge group is compact.  Since complexified $SO(3)$ is not compact,
the procedure cannot be directly applied to gravity, but some recent
work provides considerable hope that the final results can be
extended to the gravitational case.
The case of compact
gauge groups also be relevant to the study of QCD or other
Yang-Mills theories.

Ashtekar and Lewandowski [2] have studied the Ashtekar-Isham
completion \ai for the case of compact
gauge groups and have introduced both a diffeomorphism invariant
measure $\mu_{AL}$ and a differential structure on \ai.  Other diffeomorphism
invariant measures on \ai were introduced by Baez [3].
The manner in which these structures are built led to the
realization [4] that \ai can be characterized as a mathematical
structure known as a ``projective limit" constructed from
products of the gauge group.   In fact, the measure and differential
structures introduced by Ashtekar and Lewandowski are just the
projective limits of such structures on products of the gauge group.
This realization provides an important tool for studying \ai and
has the advantage that the projective limit exists even when the
group is not compact, suggesting an extension of the Ashtekar-Isham
compactification to the gravitational situation.

Returning to the case of compact gauge group, the characterization of
\ai as a projective limit allows a simple proof [4] that ${\cal A}/{\cal G}$
is a measure zero subset of \ai.  Thus, if quantum states are
taken to belong to the $L^2$ space defined by \ai and the
Ashtekar-Lewandowski measure, the actual value of a wavefunction
on the classical configuration space plays no part in
determining the state -- just as in the case of scalar quantum
field theory.  In addition, it turns out that the
Ashtekar-Lewandowski construction also defines a measure on the
original space \ag, though this measure lacks the nice property of
``$\sigma$-additivity" so that the resulting $L^2$ space on \ag
is not complete.  The completion of this space
is just $L^2($\ai$,\mu_{AL})$ [4], leading back to \ai.  This, too, is
similar to the case of scalar QFT.

Progress in studying such measures
should also contribute to an understanding of
the ``loop representation,"  which is to be related to the connection
representation by a functional integration over \ai.  Indeed, a loop
representation can now be rigorously defined as those functions of
loops that lie in the image of the ``loop transform:"
$$\Psi(\gamma) = \int_{\aim} \Psi_c(A) Tr \ H(\gamma, A) d \mu_{AL}$$
where $\Psi_c(A)$ is the connection representation of the
quantum state, $\gamma$ is a loop, and $H(\gamma, A)$ is the holonomy of the
distributional connection $A \in $ \ai around $\gamma$.  The space
\ai has the property that such holonomies are always well defined.

Another promising idea for improving the loop representation
is the introduction of so-called ``extended loops" by Gambini et. al. [5].
Each extended loop defines a gauge invariant function of connections that
can, in a certain sense, be thought of as
a ``smoothened" version of the holonomy
of connections around a loop.  That is, extended holonomies depend
on the value of the connection at all points in space, or at least on
some set of nonzero measure, as opposed to loop holonomies which
sample the connection only along a loop.  Such an extended loop
is given by a set of distributional multitensor densities that
satisfy certain conditions.
However, these multitensor fields can be expressed as linear
combinations of arbitrary smooth fields with distributional coefficients
of a well-defined form [5].

The extended holonomies suggest the definition of an ``extended loop
representation" from the connection representation through an ``extended
loop transform" in analogy with loop transform above.
The idea [7] is that quantum states
are represented as functions of extended loops and that this
is to be related to the connection representation through the
formula:
$$\Psi(M) = \int_{connections} \Psi_c(A) Tr \ W_A(M) d\mu_?$$
where $\Psi_c(A)$ is the connection representation of the state, $M$
is a set of multitensor densities that define an extended loop,
$W_A(M)$ is the extended holonomy for $M$ evaluated at the connection $A$,
and $\mu_?$ is some measure that is yet to be rigorously defined.
For many calculations in this representation,
the distributional coefficients separate from the smooth fields that
parameterize extended loops, yielding
well defined solutions and eliminating or regularizing a number of
divergences that arise in the loop representation.  In
particular, this feature of the extended loop representation
provides additional evidence that the Jones polynomial
represents a state of quantum gravity.  An additional
simplification of this representation is that only {\it linear}
functions of the multitensor densities need be considered.  This
can be seen from the extended loop transform above once it is
recognized that the extended holonomy $W_A(M)$ is linear in the
multitensor fields.

As with the connection representation,
important questions for the extended loop representation still remain.
Some of these are questions of convergence, as the extended holonomies
are given as infinite sums over
multitensor fields.  Another is
whether the extended holonomies can be defined on
the space \ai and integrated against the measures referred to above
to put the extended loop representation on a rigorous footing.
Finally, to implement the next step in the discussion of extended loops,
we must find some way
to characterize equivalence classes of extended loops under
spatial diffeomorphisms.

\vskip .5cm

[1]  A. Ashtekar and C. Isham, Class. Quantum Gravity, {\bf 9} (1992)
1433-85

[2]  A. Ashtekar and J. Lewandowski CGPG-93/8-1, to appear in
{\it Quantum Gravity and Knots} ed. by J. Baez (Oxford University Press)

[3]  J. Baez to appear in {\it Proceedings of the conference
on Quantum Topology}

[4] D. Marolf and J. Mour\~ao, in preparation

[5] C. Di Bartolo, R. Gambini, J. Griego, hep-th/9303010, to appear
in Commun. Math. Phys (1993)

[7] C. Di Bartolo, R. Gambini, J. Griego, and J. Pullin, gr-qc/9312029,
CGPG-93/12-1

\vfill
\eject

\centerline{\bf LIGO Project Report: December 1993}
\medskip
\centerline{Rochus E. Vogt, Director, LIGO Project, Caltech}
\bigskip
\bigskip
\medskip

The Nobel award to Hulse and Taylor has been a welcome development for
LIGO.  The Swedish Academy has now officially certified the existence
of gravitational waves, thus supporting LIGO's goal
to use gravitational waves in exploring the universe and for
fundamental physics.
\par
\medskip
LIGO's design parameters continue to be the frequency band from a few
Hz to several kHz, with a start-up strain sensitivity of $h \sim
10^{-21}$, and an ultimate sensitivity goal of $h \sim 10^{-23}$.\par
\medskip
Funding for LIGO R\&D and construction was \$19.1M in FY92, and \$24M
in FY93.  Congress continues to be very supportive of LIGO, and we
expect a considerable jump in funding for FY94.\par
\medskip
Land transfer for the LIGO facility at Hanford, Washington from DOE to
NSF is complete, the geotechnical site characterization is done, the
environmental FONSI (finding of no significant impact) has been
approved, and the contract for rough grading of the site is expected to
be awarded in December.  The bulldozers will start moving, weather
permitting, as soon as we have completed the obligatory ground-breaking
ceremony.\par
\medskip
Land acquisition for the Louisiana facility has been delayed by
difficult negotiations with private landowners and oil companies whose
pipelines we need to cross.  As soon as these negotiations reach a
satisfactory resolution, the (forested) land will be cleared along the
two arms of the LIGO layout, geotechnical and hydrological
characterization will be completed, and the (not at all trivial!) civil
engineering preparation of the site will commence.\par
\medskip
The engineering design contract for the 16 km of vacuum beam tubes has
been awarded to Chicago Bridge and Iron, Inc. (CBI).  The design will
be completed in March '94, to be followed by an 8-month full scale
mock-up of a partial beam tube section for verification of the
manufacturing and degassing techniques.\par
\medskip
Industrial contracts for the remainder of the vacuum system and the
building design will be placed in the first half of 1994.\par
\medskip
R\&D activities at Caltech and MIT are progressing significantly in a
variety of areas: good progress in strain sensitivity on prototypes,
the development of the design for LIGO's first interferometers
including the choice of the initial optical topology (broadband
recycled), the development of alignment systems, optical modeling, new
antivibration systems, and new mode cleaners (spatial and frequency
filters for the laser light).\par
\medskip
The 12-year old 40-m interferometer system at Caltech is being replaced
by a totally new system, called ``Mark II," which includes a much
enlarged vacuum system (2-ft diameter beam tubes and 4-ft diameter
vacuum chambers), and vastly improved ($\sim$ 3 orders of magnitude
better) vibration isolation stacks.  The interferometer is now
operating in this new system, and over the coming months its various
components will be replaced, one by one, as part of the detector R\&D.
This Mark II system, which is expected to serve us for the next decade,
will be primarily focused upon displacement noise studies, while a new
5-m system at MIT (to be activated during 1994) will be optimized for
phase noise studies. These two systems will be the key facilities for
testing and development of components for the first LIGO
interferometers. \par
\medskip
Recruitment of staff for the LIGO project has been successful, and all
key positions have been filled.  The LIGO team consists now of about 50
souls!\par
\medskip
During the past year, LIGO has also been the subject of unjustified
negative publicity.  (There also has been much positive and
enthusiastic support of LIGO --- but this tends to be less
noticed!)\par
\medskip
In particular, the LIGO team was involved in a difficult personnel
matter at Caltech that was widely publicized. Only one side of the
story was promulgated because the team, for ethical reasons, chose not
to publicize its own side.  Much of the stuff that was rumored and some
that was printed was patently untrue; we urge you not to believe
everything you heard or read.\par
\medskip
There were also mistaken assertions that LIGO funding came at the
expense of other physics support at NSF.  This is not correct.  LIGO is
an identified item in the NSF budget and funded as such by Congress.
We, in fact, fought very hard in support of the ``other physics" budget
at NSF.  There is no way that the LIGO Project would support the
funding of LIGO at the expense of individual investigator grants.\par
\medskip
LIGO will always be subject to extensive scrutiny (which is
appropriate!) and will never be safe from mistaken fears, or politics
in general.  LIGO {\bf is} the biggest project at NSF and is thus
highly visible, and is sometimes viewed with apprehension.  The LIGO
team understands that taking the heat is a price that must be paid if
one wants to be a pioneer in something very new!\par
\medskip
We appreciate the strong support that we have received from the gravity
community, and we will endeavor to keep you in touch with future
progress in LIGO.\par

\vfill
\eject

\centerline{\bf Dark Matter or New Gravity?}
\medskip
\centerline{Richard Hammond, North Dakota State University}
\bigskip
\bigskip
\medskip

There is no longer any doubt that the motion of the outer stars in galaxies
is not compatible with the gravitational theoretical predictions based on
the observed matter distribution.
Toward the outer limit of the galaxy, once nearly all of
the mass is enclosed, Newton's law predicts that the velocity is proportional
to $1\over\sqrt{r}$. However, for virtually all of the galaxies where rotation
curves (velocity {\it vs.} distance)
have been made, this does not occur. For large $r$ the velocity tends
toward a constant value, and the rotation curves become flat.

One of the original conjectures to explain this mystery was the speculation of
dark matter. Using Newtonian dynamics, it is evident that if there is
additional matter in the galaxy that is not observed {\it but behaves
gravitationally exactly like ordinary matter} then the rotation curve must be
modified. In fact, if the dark matter has a density given by
$\rho_{dark}\sim   1/r^{2}$
then it follows that, for large $r$, the gravitational force is
proportional to $1\over r$, and therefore that the velocity is constant with
distance. Thus, the postulate of the existence of dark matter with the
$1\over r^{2}$ density `solves' the problem of the flat rotation curves.

However, problems with this solution soon arose. First of all, it is
unknown why the density of dark matter should be what it is, and
further, the density profile must change for small and large $r$. Of
more immediate concern, though, is its invisibility. Typical galactic
time constants are of the order of $10^{8}yrs$. Consider hydrogen
gas. If it is in thermal equilibrium and supported by its pressure,
one may balance the gravitational attraction against the radiation
pressure. Using the ideal gas law and the presumed density for
$\rho_{dark}$, one easily shows that the temperature of this gas is of
the order of $10^{6}K$, which would be observed, but is not [1].  This
result is used to rule out the possibility that the dark matter
consists of ordinary Hydrogen.  Other arguments [2] have excluded the
possibilities that ice, dust, red and brown dwarfs could be the origin
of dark matter.  Fortunately, as fast as astrophysicists could rule
out conventional forms of matter as candidates for the dark matter,
high energy physicists have been able to elect other forms of matter
to the post. These include hot dark matter such as massive neutrinos,
cold dark matter such as WIMPS, and particle that result from specific
theories, such as axions, photinos, and so on.  In addition to new
particles, larger objects have entered the game.  Recent evidence
points to the existence of Jupiter type objects surrounding our
galaxy, dubbed MACHOS for massive compact halo objects [3].

The postulate of dark matter is not the only explanation of flat
rotation curves. Another school of thought takes aim at the use of
Newtonian dynamics.  Although general relativity (and therefore
Newtonian dynamics in the proper limit, with the proper relativistic
corrections) is experimentally verified on the scale of the solar
system, it has not undergone any test in regions larger than
that. (Cosmological solutions should not be considered as large scale
verifications due to the number of other assumptions that are
invoked.) In fact, one may argue, forsaking dark matter, that in the
only tests of gravitational theory at length scales larger then the
solar system, {\it it fails!}  Well, ever since the birth of general
relativity there have been more generalizations, modifications, and
alterations that you could shake a stick at anyway, and perhaps some
such an alternate theory is in fact responsible for the flattened
rotation curves.

The first (modern) attack on gravitational theory was launched by
Milgrom [4] who explained flat rotation curves by modifying Newton's
law of motion.  In recent years, a flash of interest, ignited by the
Fischbach conjecture, in finite range gravitational forces has been
felt.  Such a force has been derived theoretically from both quantum
gravity and classical gravity, and also appears when the dilaton field
is coupled to gravity. Sanders [5] claimed that this force accounts
for the flat rotation curves of six different galaxies, but later
other problems with this solution were uncovered [6].  Very recently,
however, this concept has been restoked by Eckhardt [7] who obtains
flat rotation curves using a two (Yukawa) potential formulation, yet
quenches the problems Sanders fired with a single potential.  From
another view, however, alternate theories may account for such
curves. In conformally invariant fourth order theory, Mannheim derives
an additional long range force, akin to that caused by the gluon field
[8] and claims to account for flat rotation curves. The color singlet
problem, or the very long range effects, of such a force are
unsettled.

The controversy between dark matter and modified gravity is not new. It
raged long ago in
trying to understand the motion of Uranus. Alternate laws of gravitation
were weighed against conjectures of unseen matter, which included
not only the prediction of
 a new planet (Neptune), but `cosmic fluid'--- the
original dark matter [9].

This brings us to the title of this article. Is the solution to
flat rotation curves the existence of dark matter or a generalized theory
of gravity?
The overall consensus in this area lies in the Dark Matter solution. By
conjuring up the appropriate distribution, it seems all motions, not only
the outer stars but galactic motions themselves, can be
satisfactorily explained. This also gives high energy physicists a lot
of business.
The character and amount of dark matter may be decisive in the choice
of the correct fundamental unified theory, if there is one.
However, if the solution lies in dark matter, and if dark matter
comes from a new or current unified or supersymmetric theory, then
this may lead to a new formulation of gravity after all.
If the dark matter
consists of a supersymmetric particle (probably the lightest supersymmetric
particle), then the rotation curves, and the inferred dark matter density,
could be the best evidence for pointing to the correct supersymmetric,
or unified, theory.

\vskip 2 cm
\noindent 1. D. J. Hegyu and K. A. Olive, Phys. Lett. {\bf 126B}, 34 (1983).

\noindent 2. {\it op. cit.}

\noindent 3. Alcock et. al., Nature {\bf 365}, 621 (1993).

\noindent 4. M. Milgrom, Astrophys. J. {\bf 270}; 365, 371, 384 (1983).

\noindent 5. R. H. Sanders, Astron. Astrophys. {\bf 154}, 135 (1986).

\noindent 6. R. H. Sanders, Mon. Nat. R. astr. Soc. {\bf 223}, 539 (1986).

\noindent 7. D. H. Eckhardt, Phys. Rev. D. {\bf 48}, 3762 (1993).

\noindent 8. P. Mannheim, Astrophys. J. {\bf 419}, 150 (1993).

\noindent 9. M. Grosser, {\it The Discovery of Neptune} (Harvard
University Press, 1962).

\vfill
\eject

\centerline{\bf Gravitational Waves from coalescing compact binaries}
\medskip
\centerline{Curt Cutler, Cornell}
\bigskip
\bigskip
\medskip

A 3-day workshop on ``Gravitational Waves from Coalescing Compact
Binaries'' was held at Caltech in early January.  Coalescing binaries
are currently viewed as the most promising source for detection by
LIGO/VIRGO. The workshop was attended by twenty-five theorists and by
several members of the LIGO experimental team.  The workshop was
organized in order to allow those theorists who are currently working
actively on this subject to compare notes and make plans for their
future research.  The central goal was to make sure that, by the time
LIGO/VIRGO goes on line (ca. 1998), all theoretical foundations have
been laid for fully utilizing the coalescing-binary observational data
that LIGO/VIRGO can bring us.

There are plans to hold, within the next year, a larger, publicly
advertised, several-day workshop on this same subject, so as to bring
other researchers up to date on it and encourage them to contribute to
the theoretical effort that will underlie LIGO/VIRGO.

Most discussions were concerned (in one way or another) with
gravitational wave templates, theoretically generated waveforms that
are used as filters to extract signals from the noisy LIGO data stream
and which are also the basis for ``reading off'' from the measured
waveform the binary's physical parameters, such as the masses of the
two bodies, their distance from Earth and angular location on the sky,
etc.  To establish a notation, we can write $$h_{GR}(t) = h_{NQ}(t) \,
+ \, {\rm post-Newtonian \ corrections}$$ where $h_{GR}(t)$ is the
full waveform predicted by general relativity and $h_{NQ}(t)$ is the
approximate version obtained using the Newtonian, quadrupole
formula. To date, the post-Newtonian corrections to $h_{NQ}(t)$ have
been calculated explicitly through $O(v/c)^3$.

\vskip .3in
The workshop focused on the research that people are currently doing
and that they intend to do, rather than on already-published
results. Accordingly, this summary will concentrate on the open
questions.  Two (related) questions which received a good deal of
attention, and which should surely yield to a little effort, are:
\vskip .1in
\noindent 1)Are the approximate waveforms $h_{NQ}(t)$
adequate as search templates
for digging signals out of the noise (even though they are clearly not
sufficiently accurate for determining the binary's parameters)?
\vskip .1in
\noindent S. Finn and D. Nicholson reported some preliminary calculations
which indicated that $h_{NQ}(t)$ is a better search template than
many of us had expected, but precisely how good it is remains to be
settled.
\vskip .2 in

\noindent 2)In practice, one will search for signals using some discrete set
of templates--how many are necessary?
\vskip .1in

\noindent The answer to question 2) is one determinant of how
much computing power
will be necessary for conducting the search.
Now, there is a natural metric on the space of waveforms,
so one might naively expect the
necessary number of search templates
to be basically the total volume of measurable waveforms
(times some number of order one which is set by the distance
between discrete search templates).  However
B. Schutz suggested that this integral probably overcounts
the total number of required templates, due to correlations between
different templates. He further suggested that there may be some
generalization of the sampling theorem which would be applicable
to this question.

\vskip .25in

Another important class of open questions relates to the accuracy with
which the binary's physical parameters (especially the masses and
spins) can be extracted from the data:
\vskip .1in

\noindent 3)What are the limitations on parameter extraction
accuracy due to random detector noise?
\vskip .2in

\noindent 4)What are the limitations on parameter extraction
accuracy due to imperfect theoretical templates?
\vskip .1in
\noindent Though much work has already been done on 3) using
successively more realistic models of the waveforms, the results have
not yet ``converged.''  The next step will be to incorporate waveform
modulation due to orbital precession caused by the bodies' spins.
Regarding 4), E. Flanagan presented preliminary results suggesting
that one will need to include post-Netownian corrections through
$O(v/c)^6$ to reach the point where systematic errors due to
inaccurate templates are smaller than random errors due to detector
noise.  This was discouraging; L. Blanchet, who is spear-heading the
PN expansion effort, expressed the opinion that carrying the expansion
through $O(v/c)^6$ would be very difficult.

All the work on 4) relies on a specific testbed for determining the
rate of convergence of the PN expansion: the problem of gravitational
radiation emitted by a point particle spiraling into a Schwarzschild
black hole.  The use of this problem as a testbed was introduced by
the Caltech group, and has been substantially advanced by recent
numerical and analytical calculations by H. Tagoshi, T. Nakamura, and
M. Sasaki.  These calculations were presented at the workshop by
Tagoshi, and Flanagan's $O(v/c)^6$ estimate is based on them.

\vskip .25in

The workshop also had a section on waveforms from the final merger of
neutron-star binaries, and on the possible LIGO/VIRGO use of those
waveforms to measure neutron-star radii.  J. Centrella, M. Davies, and
T. Nakamura all presented their recent numerical simulations of such
mergers, and D. Laurence displayed and compared spectra ($dE/df$) of
the gravitational waves from the various simulations. The most
interesting results were (i) that the mergers were qualitatively
different depending on whether or not the neutron stars were assumed
to be rapidly rotating just prior to collision, with much more
post-collision ``ringing'' in the non-rotating case, and (ii) in all
cases there is a rather sharp cliff in the spectrum which is quite
sensitive to the neutron-star radii.  A. Wiseman warned, however, that
PN effects, not-yet included in the simulations, may reduce
substantially the sharpness of the cliff and its viability as a
measure of neutron-star radii.  \vskip .25in

Many other topics were discussed, but the above should give a flavor
of what went on at the workshop.  Most participants seemed to find the
experience very stimulating.  Roughly $40 \%$ of the participants
expressed strong interest in attending a much longer workshop--one
that would last for several months.
\vskip .1in
Finally, since any interest group these days must have an electronic
bulletin board, a merging-binaries bulletin board was established for
us by S. Finn.  Those wishing information about the bulletin board, or
wishing to be placed on the associated mailing list, should send the
request to

\noindent gwave-theory-request@holmes.astro.nwu.edu.

\vfill
\eject

\centerline{\bf Mach's Principle: From Newton's Bucket to Quantum Gravity}
\medskip
\centerline{Dieter Brill, University of Maryland}
\bigskip
\bigskip
\medskip

An international workshop on Mach's Principle was held July 26 - 30,
1993 at T\"ubingen, Germany . It was the first meeting ever devoted
exclusively to this elusive principle, and was attended by about 50
physicists, a diverse group that fairly represented the various views
on the Principle extant today.  The organizers were J. Barbour
(Oxford) and H. Pfister (T\"ubingen), with an advisory committee of
B. Bertotti, D. Brill, J. Ehlers and J. Stachel.

It was the aim of the conference to give a reasonably complete
overview of the activities derived from Mach's original questions,
ranging from history of physics to general relativity theory and to
observational astronomy. Recognizing the variety of views, J. Barbour
suggested polling the conferees at the beginning and the end on the
proposition that ``Einstein's general relativity is a perfectly
Machian theory". (The votes gave no decisive indication of any strong
change of minds by the end of the conference; the majority opined that
GR, even with ``appropriate boundary conditions of closure of some
kind" is not perfectly Machian, but agreed --- overwhelmingly at the
end --- that GR is ``very Machian.")

A number of papers were devoted to ways of giving a precise
formulation of Mach's principle. A random sampling from this group
follows.  J. Narlikar and F. Hoyle spoke about direct particle
formulations based on action at a distance similar to Wheeler-Feynman
electrodynamics. There is no solution describing an empty universe,
and the smallest possible number of particles appears, amusingly, to
be three.  In the limit of a large number of particles the theory
resembles general relativity, but also allows interpretation in terms
of creation of matter.  D. Raine and K.  Nordvedt presented
contributions on the Green's function formulation.  An integral
representation formulates perhaps most directly the idea that the
universe's matter content should determine the local inertial
frames. The Discussion included questions about the epoch at which the
integral representation is to be applied, and whether ``true" matter
should be distinguished from other sources of gravity, such as black
holes. J. Isenberg reported on the ``Wheeler-Einstein-Mach"
formulation in terms of initial values. Here the emphasis is to
characterize which features of gravity are freely disposable, and
which are thereby determined. But some of the freely disposable
features needed to determine inertia would seem unphysical, because
hidden behind horizons. The question whether alternative schemes are
needed or whether general relativity itself is perfectly Machian was
discussed by J. Barbour, D. Lyndon-Bell, and others.

The best-understood Machian effect in general relativity and related
theories is the dragging of inertial frames by rotating bodies.
In addition to the rather general but perturbative calculations,
reviewed by H. Pfister, there are now exact solutions for a rigidly
rotating disk of dust, as presented by R. Meinl. Experimental tests
of such Machian effects were discussed by C. Will, I. Ciufolini, and
others.

In addition to the expository talks there were contributions
reviewing critically various fundamental issues, by J. Ehlers, W. Rindler,
H. Bondi and others. The meeting was rounded out by numerous spirited
discussions, for which there was fortunately enough time. Only the
discussion of Mach's principle in quantum gravity, at the very
end of the sessions, failed to do justice to this interesting topic.

A number of the T\"ubingen meeting's pleasant and interesting arrangements
deserve mention. It was held at the Max Planck House of the local
Max Planck Institute, a nearly ideal location for a small conference
that fosters lively interaction between the participants. On the
evening that had the week's most pleasant weather there was a party
at the home of Prof.~Pfister that will remain memorable for its
delicacies, both intellectual and gustatory. Equally memorable was
a visit to a jewish cemetery at Bad Buchau that contains many graves
of Einstein's forefathers.

In his closing remarks, Sir Hermann Bondi said: ``This conference was
a splendid idea, and I am only surprised that nobody thought of
having such a conference before."

Many of the contributions to the T\"ubingen conference will appear in
the {\it Einstein Studies} series published by Birkh\"auser.

\vfill
\eject

\centerline{\bf Cornelius Lanczos International Centenary Conference}
\medskip
\centerline{David Brown, North Carolina State University}
\bigskip
\bigskip
\medskip

The Cornelius Lanczos International Centenary Conference was sponsored
by the College of Physical and Mathematical Sciences of North Carolina
State University, and took place in Raleigh on December 12--17,
1993. The conference was held in honor of Cornelius Lanczos, who
contributed important and fertile ideas in an astonishingly broad
range of disciplines. Lanczos pioneered the study of
higher--derivative gravity theories and the use of Euclidean methods
in relativity.  His early work was instrumental in revealing the
non-singular nature of event horizons and in uncovering their physical
properties. In a 1926 paper, Lanczos proposed a field formulation of
quantum mechanics using integral equations, in anticipation of
Schr\"odinger's equation. Lanczos also made important contributions to
electromagnetic theory, and his book entitled {\it Variational
Principles of Mechanics\/} is a testimony to his mastery of that
subject. Lanczos's contributions to applied mathematics are likewise
impressive. He laid the foundation for the Fast Fourier Transform, and
developed a number of other computational algorithms that are
currently of vital importance in the field of applied mathematics.

The topics covered at the Lanczos Conference included theoretical
physics and computational mathematics, reflecting the scientific
interests of Cornelius Lanczos. As a modern bridge between these
disciplines, the conference also covered various aspects of
astrophysics. A number of joint sessions brought together scientists
{}from each of these fields. The joint plenary speakers included Roger
Penrose, who spoke on ``Relativity, Quantum Theory, and Computation".
Penrose discussed the distinction between deterministic systems and
computable systems, and suggested that our Universe might not be
computable even if it is deterministic. John Stachel gave an
interesting and informative talk on ``Lanczos's Contributions to
General Relativity". Kip Thorne's presentation, ``Gravitational Waves:
Challenges, Plans and Prospects", provided a summary of the
astrophysical information carried by gravitational waves, the
prospects for detecting and extracting information from gravitational
waves, and the challenges that lie ahead. In his talk entitled ``The
Quasiclassical Domain in a Quantum Universe", James Hartle discussed
the idea that the quasiclassical domain originates from the
fundamental interactions along with the specific initial conditions of
our universe.  Hartle went on to review his work with Murray
Gell--Mann on the derivation of deterministic physics through the
decoherent histories generalization of quantum mechanics.

The joint sessions also featured several speakers from the
computational mathematics community. Gene Golub discussed various uses
for the `Lanczos algorithm', which Lanczos originally developed as a
method for computing the eigenvalues and eigenvectors of
matrices. Beresford Parlett's talk addressed some subtle difficulties
with the Lanczos algorithm and their resolution. James Cooley lectured
on the work of Lanczos that contains the essential ideas of the Fast
Fourier Transform, and covered the relationship between Lanczos's work
and the currently familiar formulations of the FFT. Anne Greenbaum
discussed the finite precision implementation of the Lanczos algorithm
and stressed the benefits of this approach.

The remainder of the conference featured concurrent programs in
computational mathematics and in theoretical physics and
astrophysics. A highlight of the theoretical physics and astrophysics
program was the lecture ``Towards `It from Bit'" by John Wheeler,
which contained a wealth of insights and observations. Abhay Ashtekar
reported on recent progress in developing a theory of integration on
the space of connections modulo gauge transformations. This and other
related work was discussed in the minisymposium ``New Variables and
Loop Quantization" organized by Lee Smolin. Karel Kucha\v{r} organized
a minisymposium that addressed ``The Problem of Time in Quantum
Gravity", while Jonathan Halliwell's minisymposium on ``Decoherence
and the Foundations of Quantum Mechanics" provided further discussion
of the issues raised in Hartle's talk.

Tsvi Piran presented evidence in his plenary talk that the observed
$\gamma$--ray bursts originate in the merger of neutron star
binaries. Michael Turner discussed the problem of explaining in detail
how the small, primordial fluctuations in matter density lead to the
presently observed structure in our universe.  This and other related
topics were discussed in the minisymposium ``Galaxy Formation and
Large--Scale Structure of the Universe" organized by Alex
Kashlinsky. Clifford Will's minisymposium ``Detection of Gravitational
Radiation from Astrophysical Sources" served as a complement to
Thorne's plenary lecture. The minisymposium organized by Manfred
Scholer and Dan Winske covered ``Numerical Simulations of
Collisionless Space Plasmas", while the minisymposium organized by
John Blondin and James Stone focused on ``Computational
Magnetohydrodynamics in Astrophysics".

Several of the plenary speakers chose black holes as the topic of
their lectures.  Claudio Teitelboim discussed the fundamental role
played by the Gauss--Bonnet theorem in determining the entropy of a
black hole through Euclidean path integral methods.  Robert Wald
presented a technique, based on variational principles, that yields a
generalized first law of black hole mechanics for essentially any
diffeomorphism invariant Lagrangian field theory. The lecture by Gary
Horowitz concerned pair creation of magnetically charged black holes
by magnetic fields. His results suggest that the rate of creation of
extremal black holes is suppressed relative to the rate of creation of
spatial wormholes by the exponential of the Bekenstein--Hawking
entropy. These talks were complemented by the minisymposium ``Black
Hole Evaporation and Thermodynamics" organized by Paul Anderson.

Several minisymposia were devoted to important problems in classical
general relativity. These included a session on ``Cosmic Censorship"
organized by David Garfinkle, and a session on the ``Lanczos
$H$--tensor" organized by Patrick Dolan and Abraham Taub. In the early
1920's, Lanczos addressed the Cauchy problem of general relativity and
showed that it is well posed. The ongoing efforts to understand this
problem were discussed in the minisymposium ``The Cauchy Problem of
General Relativity" organized by Jim Isenberg.

The theoretical physics and astrophysics program also included a
plenary talk by Jerrold Marsden, who described a general method for
generating a constrained variational principle governing the
Euler--Poincare equations on any Lie algebra. Complementary
minisymposia included ``Geometric Mechanics" organized by Tony Bloch
and Tudor Ratiu, and ``Symplectic Methods in Physics" organized by
Mark Gotay and Peter Olver.

Theoretical high energy particle physics was represented at the
Lanczos Conference in the minisymposia ``Supercollider Physics"
organized by Paul Frampton, Tom Kephart, and Marc Sher, and ``Open
Questions in Particle Theory" organized by Carl Carlson and Adam
Szczepaniak. Unfortunately, Yasushi Takahashi was unable to attend and
deliver his plenary lecture ``Four Dimensional Vector and the Gauge
Transformation" due to a cancelled airline flight.

In addition to the plenary talks and minisymposium sessions, the
Lanczos Conference featured a large number of contributed lectures and
poster presentations covering a wide spectrum of topics in theoretical
physics, astrophysics, and computational mathematics. A special public
lecture by Michael Turner entitled ``The Earliest History of the
Universe" was a great success, attracting approximately 700 people
{}from the Raleigh community. The week--long conference was fun and
informative for the nearly 600 participants. Proceedings of the
Lanczos Conference will be published in the summer or fall of 1994.

\vfill
\eject

\centerline{\bf  Third Midwest Relativity Conference }
\medskip
\centerline{David Garfinkle, Oakland University}
\bigskip
\bigskip
\medskip

The third annual midwest relativity conference was held at Oakland
University in Rochester Michigan on Nov. 5-6.  There were about 50
participants, most of whom gave talks.  Each talk was 15 minutes long.
For the most part the topics covered fell into the following
categories: 1) Black Holes, 2) Numerical relativity 3) Mathematical
relativity 4) Observational topics 5) Quantum gravity.

{\bf Black Holes.} Many of the talks on black holes concentrated on
the notion of black hole entropy, both in general relativity and in
other theories of gravity.  Bob Wald presented a general method for
finding the entropy of a stationary black hole in any diffeomorphism
invariant theory.  Vivek Iyer and Ted Jacobson presented some of the
results of this method and addressed the issue of extending the method
to the case where the black hole is not stationary.  Robert Meyers
considered the case of $R^2$ gravity and found conditions under which
this theory obeys the second law of thermodynamics.  David Garfinkle
showed that there is a relation between black hole entropy and the
rate of production of black holes by quantum tunneling.  Gungwon Kang
presented a conformally invariant generalization of the surface
gravity of black hole.  Jonathan Simon considered the issue of
boundary terms in a Euclidean formulation of higher derivative
gravity.  Robert Mann presented some exact solutions for two
dimensional Liouville black holes.  Mike Morris showed why the
Kerr-Newman metric is not a good model for an elementary particle.
\vskip0.2truein

{\bf Numerical Relativity.} Numerical results for problems in 1, 2
and 3 spatial dimensions were presented; some in vacuum and some using
matter.  Ed Seidel and Wai-Mo Suen presented results for 2 and 3
dimensional codes describing the evolution of a distorted black hole.
They also discussed their method of ``horizon locking'' which allows
the numerical grid to end on the apparent horizon.  Beverly Berger and
Vijaya Swamy presented a numerical evolution of Gowdy spacetime.  They
investigated the interactions of gravity waves both with each other
and with test matter.  Comer Duncan presented a numerical algorithm
for solving Poisson's equation in 3 dimensions and accurately finding
the gradient of the solution.  Malcolm Tobias presented a numerical
evolution of plane gravitational waves as a code test of the 3
dimensional numerical code discussed by Seidel and Suen.  Jayashree
Balakrishna presented a numerical evolution of boson stars.  Daniel
Holz numerically searched for closed trapped surfaces in the initial
data for Brill waves.
\vskip0.2truein

{\bf Mathematical Relativity.} The mathematical talks were mostly
about global issues such as singularities, topology and the concept of
mass in general relativity.  John Friedman presented a proof that in
an asymptotically flat spacetime exotic topology is hidden behind an
event horizon and cannot be seen by an observer at infinity.  Ed Glass
and Mark Naber presented results on Taub Numbers.  These are
quantities associated with perturbations of a background spacetime
that has a Killing vector.  John Beem showed that if a spacetime is
singular then, under certain conditions, all ``nearby'' spacetimes are
singular.  Steve Harris showed that in a static, globally hyperbolic,
nonsingular spacetime the curvature must satisfy certain fall off
conditions.  Jim Wheeler classified all the types of singularities
possible in a spherically symmetric spacetime.  Shyan-Ming Perng used
spinorial techniques to define an analogue of mass for any initial
data set (whether or not it is asymptotically flat).  Jolien Creighton
analyzed some of the properties of the Brown-York quasilocal mass.
Matt Visser showed that the techniques of Lorentz geometry can be
applied to the problem of perturbations of a fluid.  Brien Nolan
applied Israel's thin shell formalism to model a shell of matter in a
Robertson-Walker spacetime.  Jim Chan presented exact solutions of a
generalization of 2 dimensional dilaton gravity.
\vskip0.2truein

{\bf Observational Topics.} The observational talks were mostly
related to LIGO, its uses and possible sources.  Sam Finn talked about
using LIGO to measure the Hubble parameter.  Robert Caldwell discussed
a stochastic background of gravity waves from inflation and the
possibility of observing these gravity waves using the cosmic
microwave anisotropy.  Alan Wiseman and Lawrence Kidder used the post
Newtonian approximation to treat the case of coalescing binaries.
Liliana Simone treated head on collisions between black holes in the
post Newtonian approximation.
\vskip0.2truein

{\bf Quantum Gravity.} There are many approaches to quantum gravity;
some (though not all) were presented at this meeting.  Dieter Brill
talked about using instanton methods to calculate the rates for
topology changing quantum tunneling processes.  Ian Redmount used the
minisuperspace approach to treat the quantum dynamics of spherical
wormholes.  Jorma Louko presented some results on the quantization of
2+1 dimensional gravity on the manifold $ R \times {T^2}$.  Miguel
Ortiz talked about the issue of physical states in $ N=1$
supergravity.  Gilad Lifschytz did quantum field theory on the curved
space of a 2+1 dimensional string theory black hole.  Greg Daues did
quantum field theory on the curved space of a Brans-Dicke type
cosmology.  Charles Torre discussed the problem of observables in
canonical quantum gravity.  Michael Reisenberger studied the loop
representation of abelian gauge theories.

\vskip0.1truein

There was a wide range of research in general relativity presented at
this meeting, as at the two previous midwest relativity conferences.
I hope to see
all of you at the next midwest conference in St. Louis.

\end